# Spatial Auditory Brain-computer Interface using Head Related Impulse Response

Chisaki Nakaizumi[1,*], Toshie Matsui[1], Koichi Mori[2], Shoji Makino[1], and Tomasz M. Rutkowski[1,3]
[1] Life Science Center of TARA, University of Tsukuba, Japan
[2] Research Institute of National Rehabilitation Center for Persons with Disabilities, Japan
[3] RIKEN Brain Science Institute, Japan
e-mail: tomek@bci-lab.info & http://bci-lab.info/

*Abstract* This study reports on a head related impulse response (HRIR) application to an auditory spatial brain-computer interface (BCI) speller paradigm. Six experienced and five BCI-naive users participated in an experimental spelling set up based on five Japanese vowels. Obtained auditory evoked potentials resulted with encouragingly good and stable P300-responses in online BCI experiments. Our case study indicated that the headphone reproduced auditory (HRIR-based) spatial sound paradigm could be a viable alternative to the established multi-loudspeaker surround sound BCI-speller applications, as far as healthy pilot study users are concerned.

*Keywords*  Auditory BCI; P300; EEG; brain signal processing.

## I. INTRODUCTION

BCI is expected to provide a speller for paralyzed people such as those suffering from an amyotrophic lateral sclerosis (ALS), etc. The majority of successful BCI applications rely on motor imagery or visual modalities, which require long-term training and good eyesight from the users [1]. Auditory modality could be an alternative method because it does not require good eyesight, although it is still not as successful as the visual applications. We propose to extend the previously reported vector-based-amplitude-panning (VBAP)-based spatial auditory BCI (saBCI) paradigm [2] by making use of a head-related-impulse-response (HRIR) [3] for virtual sound images spatialization with headphones-based sound reproduction. In addition we simplify the previously reported real sound sources generated with surround sound loudspeakers [4]. The HRIR appends interaural-intensity-differences (IID), interaural-time-differences (ITD), and spectral modifications to create the spatial stimuli. The HRIR allows also for a more precise and fully spatial virtual sound images positioning utilizing even not the user own head's HRIR measurements.

## II. METHODS

All of the experiments were performed in the Life Science Center of TARA, University of Tsukuba, Japan. We conducted pilot experiment with the six BCI experienced users and a practical previous study with BCI-naive participants [5]. We chose five Japanese vowels for the first step for the BCI-speller. We spatialized the sound stimuli using a public domain CIPIC HRTF database [3]. Each Japanese vowel was set spatially on a horizontal plane at azimuth locations of -80°, -40°, 0°, 40°, and 80° for the *a, i, u, e,* and *o* vowels, respectively. Synthetic voice was used in the pilot and recorded voice [6] in the practical experiments, respectively. The online EEG experiments were conducted to investigate P300 response validity. The brain signals were collected by a bio-signal amplifier system g.USBamp by g.tec Medical Engineering GmbH, Austria. The EEG signals were captured by sixteen active gel-based electrodes attached to the following head locations *Cz, Pz, P3, P4, Cp5, Cp6, P1, P2, Poz, C1, C2, FC1, FC2,* and *FCz* as in the extended 10/10 international system. The ground electrode was attached on the forehead at the *FPz* location, and the reference on the user's left earlobe, respectively. The extended BCI2000 software was used for the saBCI experiments to present stimuli and display online classification results. A single experiment was comprised of five sessions and a session contained of five selections. In brief, single experiment was comprised of 25 selections. A single selection was comprised of 10 *targets* and 40 *non-targets*. Brain signals were averaged 10 times for each vowel classification. The single stimulus duration was set to 150 ms in pilot experiment and 250 ms in practical experiment [5]. The inter-stimulus-interval (ISI) was set to 350 ms in the pilot and 150 ms in practical experiment. The EEG sampling rate was set to 512 Hz and the notch filter to remove electric power lines interface of 50 Hz was applied in a rejection band of 48 to 52 Hz. The band pass filter was set with 0.1 Hz and 60 Hz cutoff frequencies. The acquired EEG brain signals were classified online within the in-house extended BCI2000 application using a stepwise linear discriminant analysis (SWLDA) classifier with features drawn from the 0~800 ms ERP intervals.

## III. RESULTS AND DISUCUSSIONS

This section presents and discusses results obtained from the EEG experiments conducted with eleven users as described in the previous sections. The average and the best accuracies are summarized in **Table 1** with a major comparison measure called an information-transfer-rate (ITR). Users #E1~E6 were the BCI experienced of the pilot experiment and #N1~N5 the naive from the practical sessions [5].

Table 1. Accuracy and ITR of all EEG experiment

| User | Averaged | The best | ITR |
|---|---|---|---|
| #E1 | 55% | 80% | 4.8 |
| #E2 | 50% | 80% | 4.8 |
| #E3 | 65% | 100% | 9.29 |
| #E4 | 60% | 100% | 9.29 |
| #E5 | 50% | 80% | 4.8 |
| #E6 | 55% | 80% | 4.8 |
| #N1 | 48% | 80% | 9.6 |
| #N2 | 64% | 100% | 18.58 |
| #N3 | 32% | 80% | 9.6 |
| #N4 | 24% | 40% | 1.21 |
| #N5 | 44% | 80% | 9.6 |

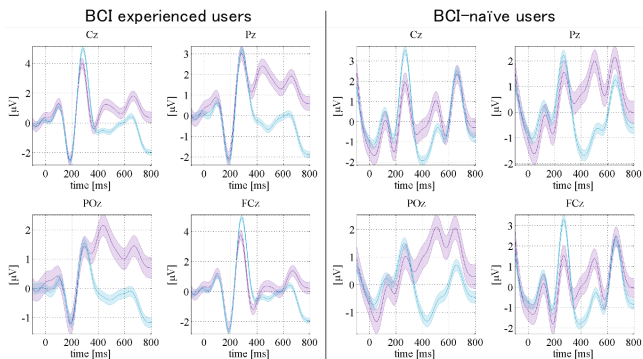

Figure 1. Grand mean average ERP brain responses.

The grand mean averaged brain ERP responses at representative four center electrodes from the both experiments are presented in **Figure 1**. The purple lines depict the brainwave responses to *targets*, and blue to the *non-targets*. The clear P300 responses have been confirmed in the both conducted experiments. The averaged BCI classification accuracies were still not perfect owing too short ISIs set in the practical experiment, yet the ITR scores improved.

## IV. CONCLUSIONS

The presented EEG results confirmed P300 reponses not only of the BCI experienced users but also in case of the BCI-naive group. The mean accuracy was not perfect yet, but the ITR results were encouraging for the novel saBCI modality. We plan to expand this modality with more letters using additionally various spatial sound elevations.

### ACKNOWLEDGMENTS
This research was supported in part by the SCOPE grant no. 121803027 of The Ministry of Internal Affairs and Communication in Japan.